\documentclass[12pt,preprint]{aastex}

\newcommand{\myemail}{jpbrown@head.cfa.harvard.edu}

\newcommand{\cxc}{{\sl Chandra X-ray Center}}
\newcommand{\cha}{{\sl Chandra}}
\newcommand{\chandra}{{\sl Chandra}}
\newcommand{\rosat}{\textit{ROSAT}}
\newcommand{\fuse}{\textit{FUSE}}
\newcommand{\bet}{Betelgeuse}
\newcommand{\ctrt}{ct~s$^{-1}$}
\newcommand{\lx}{L_{\rm X}}
\newcommand{\lbol}{L_{\rm bol}}
\newcommand{\fx}{f_{\rm X}}
\newcommand{\surffx}{F_{\rm X}}

\shorttitle{\chandra\ Observations of Betelgeuse}
\shortauthors{Posson-Brown et al.}

\begin{document}

\title{Dark Supergiant: \chandra's Limits on X-rays from \bet}
\author{
Jennifer Posson-Brown,
Vinay L. Kashyap,
Deron O. Pease,
and Jeremy J. Drake
}
\affil{Smithsonian Astrophysical Observatory \\ 
60 Garden Street, MS-21 \\ Cambridge, MA 02138}
\email{\myemail}
\email{vkashyap@cfa.harvard.edu}
\email{dpease@cfa.harvard.edu}
\email{jdrake@cfa.harvard.edu}

\begin{abstract}
We have analyzed \chandra\ calibration observations of \bet\ ($\alpha$~Ori,
M2\,Iab, m$_V=0.58$, 131~pc) obtained at the aimpoint
locations of the HRC-I (8 ks), HRC-S (8 ks), and ACIS-I (5 ks).  \bet\
is undetected in all the individual observations as well as
cumulatively.  We derive upper limits to the X-ray count
rates and compute the corresponding X-ray flux and luminosity upper limits for
coronal plasma that may potentially exist in the atmosphere of \bet\
over a range of temperatures, $T=0.3-10$~MK.
We place a flux limit at the telescope of
$\fx\approx4\times10^{-15}$~ergs~s$^{-1}$~cm$^{-2}$ at $T=1$~MK.  The
upper limit is lowered by a factor of $\approx3$ at higher
temperatures, roughly an order of magnitude lower than that
obtained previously.  Assuming that the entire
stellar surface is active, these fluxes correspond to a surface flux
limit that ranges from $30-7000$~ergs~s$^{-1}$~cm$^{-2}$ at $T=1$~MK,
to $\approx 1$~ergs~s$^{-1}$~cm$^{-2}$ at higher temperatures, five
orders of magnitude below the quiet Sun X-ray surface flux.  We
discuss the implications of our analysis in the context of models of a
buried corona and a pervasive magnetic carpet.  We rule out the
existence of a solar-like corona on \bet, but cannot rule out the
presence of low-level emission on the scale of coronal holes.

\end{abstract}


\section{Introduction}
\label{s:intro}

\begin{deluxetable}{lll}
\tablecolumns{2}
\tabletypesize{\small}
\tablecaption{Stellar parameters for \bet\ \label{t:stelpar}}
\tablewidth{0pt}
\startdata
\hline
\hline
Other Names & \multicolumn{2}{l}{$\alpha$~Ori / 58 Ori / HD~39801 / HR~2061 / SAO 113271 / HIP 27989 } \\ 
(R.A., Dec) & (05:55:10.3053,~+07:24:25.426) & ICRS 2000.0 \\
($l_{\rm II}, b_{\rm II}$) & ($199^{\circ}.79, -8^{\circ}.96$) & SIMBAD \\
Spectral Type & M2~Iab & SIMBAD \\
$m_{V}$ & 0$^m$.58 & SIMBAD \\
$B - V$ & 1$^m$.77 & SIMBAD \\
T$_{\rm eff}$ & 3650~K & Levesque et al.\ (2005) \\
$B.C.$ & -1.6 & Levesque et al.\ (2005) \\
Distance & $131 \pm 28$ pc & Hipparcos (Perryman et al.\ 1997) \\ 
$\lbol$ & $1.2~\times~10^{38}$~ergs~s$^{-1}$ & \hfil \\
Angular Diameter & $44.6 \pm 0.2$ mas & CHARM (Richichi \& Percheron 2002) \\
Radius & $631 \pm 134$ R$_{\odot}$ & \hfil \\
Mass & $14$ M$_{\odot}$ & Figure~\ref{f:evoltrack} \\
Age & $10$ Myr & Figure~\ref{f:evoltrack} \\
Gravity & $1\pm0.4$~cm~s$^{-2}$ & \hfil \\
Circumstellar Column & $6.2 \times 10^{21}$~cm$^{-2}$ & Hagen (1978) \\
\enddata
\end{deluxetable}

\begin{figure}[htb!]
\begin{center}
{\includegraphics{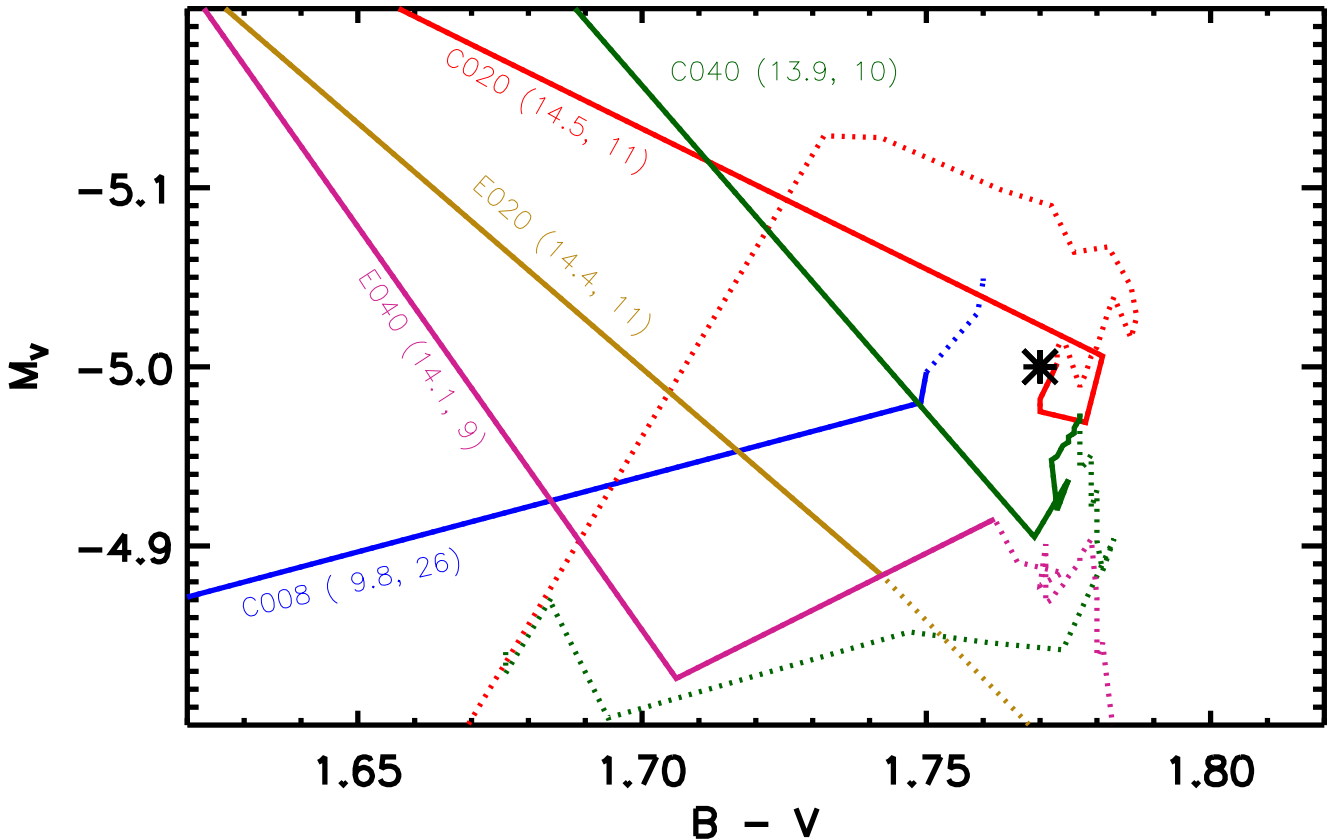}}
\caption{Evolutionary state of \bet.  The tracks from the Geneva
stellar evolutionary models (Lejeune \& Schaerer 2001) are overlaid
on a color-magnitude diagram, with the position of \bet\ represented
by an asterisk.  The lines are solid until their closest approach
to \bet\ in their evolutionary history, after which they are depicted
as dotted lines (note that the magnitude range is different for the
two axes).  The lines are labeled by the model used (the models
differ in their masses, mass-loss, and metallicities;
see Lejeune \& Schaerer for a detailed description). The mass (in M$_{\odot}$) and age (in Myr) of the star at the
point of closest approach is also shown for each model.
Inspection of the figure shows
that while definitive measurements of the mass and age of \bet\ are
not feasible due to the large systematic model uncertainties (e.g., masses ranging
from 10 to 15 M$_{\odot}$ are plausible) but the majority of the models
correspond to a mass of $\approx14$~M$_{\odot}$ and an age of $\approx10$~Myr,
which we then adopt as the nominal mass and age of \bet\ throughout this
paper.
\label{f:evoltrack} }
\end{center}
\end{figure}

\bet\ (see Table~\ref{t:stelpar}) is a nearby (131~pc), bright ($V=0.58^m$)
evolved
red supergiant star (M2\,Iab, $B-V=1.77$).
It has been monitored extensively in the optical (e.g., Wilson et
al. 1997, Burns et al. 1997), and
displays irregular brightness variations ($V=0.3-0.8$, Gray 2000) that
have been interpreted as large-scale surface structures or activity
(e.g., Lim et al.\ 1998 and Gray 2000).  
A definitive estimate of its
age does not exist since it cannot be identified with nearby
Orion associations (Lesh 1968), and estimates of its mass vary from
$5$~M$_{\odot}$ (Dorch 2004) to $10-30$~M$_{\odot}$ (Gray 2000; also
Lambert et al.\ 1984 and references therein).
However, reasonable estimates can be obtained by comparing the
evolutionary tracks of high-mass stars with the location of \bet\
on a color-magnitude diagram.  We show such a comparison using
the Geneva stellar evolutionary tracks (Lejeune \& Schaerer 2001)
in Figure~\ref{f:evoltrack}.
The stellar models vary in initial mass, metallicity, and mass-loss
rates, and no rotation effects are included.  While the systematic
model uncertainties prevent a definitive assessment of the
evolutionary history of \bet, note nevertheless that
most of the models predict similar values of age and mass for
the star, and therefore an approximate estimate
of the gross properties of \bet\ is possible.
Based on Figure~\ref{f:evoltrack}, we adopt a current
mass of $\approx14$~M$_{\odot}$ and an age of $\approx10$~Myr
for \bet.  The uncertainty in these parameters do not affect
our conclusions.

\bet\ has never been detected in X-rays, and is in a region of the
H-R diagram where a stable corona is not expected to exist
(Ayres et al.\ 1981, Linsky \& Haisch 1979, Haisch, Schmitt, \&
Rosso 1991, Rosner et al.\ 1995, H\"{u}nsch \& Schr\"{o}der 1996,
H\"{u}nsch et al.\ 1998).
Nevertheless, there
is evidence from numerical MHD simulations that dynamo action
which can produce large scale magnetic fields can exist on such
stars (Dorch 2004).  Furthermore, chromospheric activity
suggesting the presence of coronae has been detected in numerous
late-type giant stars: coronal proxy lines such as Si\,IV and
C\,IV have been detected in ostensibly non-coronal giants such
as Arcturus and Aldebaran (Ayres et al.\ 2003); and forbidden
coronal lines have been reported in \fuse\ observations
of $\beta$ Cet by Redfield et al.\ (2003).  Thus, while unlikely,
it is plausible that hot coronal plasma may exist on supergiants
like \bet.

Here we present an analysis of X-ray observations of \bet\
obtained as part of the science
instrument calibration program of the \chandra\ X-Ray Observatory (\chandra).
The data are described in \S\ref{s:data}.
\bet\ is not detected in X-rays in any of our individual observations
or in co-added data, and in \S\ref{s:ul} we set the most stringent
upper limits to the X-ray flux determined thus far, improving upon
limits obtained from the \rosat\ All-Sky Survey (RASS) by two orders
of magnitude.  In \S\ref{s:discuss}, we discuss our results
in the context of models of coronae on late-type stars.
We summarize in \S\ref{s:summary}.

\section{Data}
\label{s:data}

\subsection{\chandra }
\label{s:data_chandra}

\bet\ was periodically observed by \chandra\ from 2001 till 2007
as part of a calibration program to monitor
the UV, optical, and IR response of all of the detectors.
Of these, we consider the on-axis observations with the
ACIS-I, HRC-I, and HRC-S detectors, for which the expected
out-of-band contamination is vanishingly small and thus
can be ignored (Wolk 2002 for ACIS-I; and
Posson-Brown \& Kashyap 2005 for HRC,
see Table~\ref{t:uvtab}).  
The out-of-band leak is large for the ACIS-S detector, so we do not use any \bet\
ACIS-S observations in this analysis.
A total of $\sim 21$ ks of exposure has been accumulated
thus far with \chandra\ (see Table~\ref{t:obsinfo}).

\begin{deluxetable}{cccc}
\tabletypesize{\small}
\tablecaption{X-ray Observations of \bet\ \label{t:obsinfo}}
\tablewidth{0pt}
\tablehead{
\colhead{Obs ID} & \colhead{Instrument} & \colhead{Date} &
\colhead{Exposure (s)} 
}
\startdata
RASS & \rosat/PSPC, bands 1-7 & 1990-07-30\tablenotemark{a} &
460\tablenotemark{b} \\
RASS & \rosat/PSPC, bands 1-2 & 1990-07-30\tablenotemark{a} &
460\tablenotemark{b} \\ 
\\
3365 & \cha/ACIS-I & 2001-12-16 & 4897.2 \\ 
\\
2595 & \cha/HRC-I & 2001-12-07 & 1892.1 \\
3680 & \cha/HRC-I & 2003-02-06 & 1893.4 \\
5055 & \cha/HRC-I & 2004-02-02 & 2075.9 \\
5970 & \cha/HRC-I & 2005-02-02 & 2129.4 \\ 
\\
2596 & \cha/HRC-S & 2001-12-07 & 1926.7 \\
3681 & \cha/HRC-S & 2003-02-06 & 1819.6 \\
5056 & \cha/HRC-S & 2004-02-02 & 1945.3 \\
5971 & \cha/HRC-S & 2005-02-02 & 2140.4 \\ 
\enddata
\tablenotetext{a}{Start date of RASS}
\tablenotetext{b}{Estimated from nearby sources (see \S\ref{s:data})}
\end{deluxetable}

We use level 2 photon event lists downloaded from the public archive, which
have gone through the standard pipeline processing by the CXC
software (ASCDS versions 6.4 and above).  Further analysis was done using
CIAO software (v3.2.2) and custom
IDL\footnote{Interactive Data Language, ITT Visual Information Solutions}
software, including routines from the Package for Interactive Analysis
of Line Emission (PINTofALE; Kashyap \& Drake
2000).\footnote{Available from http://hea-www.harvard.edu/PINTofALE/}  

For both HRC detectors, we used circular source regions with $r\approx1''$,
enclosing 95\% of the X-ray PSF, centered on the location of the
source (verified with ACIS-S data, where \bet\ is visible due to the
optical leak).  To measure the background, in the
HRC-I we used annuli $30''$ wide, beginning $1'$ away from the source
location; for the HRC-S, we used annuli $45''$ wide, beginning $35''$
away from the source location.

For ACIS-I, we first followed CIAO threads to reprocess the event
list from level 1, removing afterglow events and rejecting events with
bad grades.  We then filtered the new level 2 event list to include
PI$=20-170$,
which corresponds to an energy cut of approximately $0.3-2.5$~keV.
For the source region, we used a circle with $r\approx3''$, enclosing
99\% of the ACIS X-ray PSF, and a background annulus $90''$ wide,
beginning $10''$ away from the source location.  To find the combined
\chandra\ counts upper limit, we used circular source regions with
$r=10''$ and annular background regions with $r_{min}=10''$ and
$r_{max}=100''$ for all three detectors (HRC-I, HRC-S, and ACIS-I).
The sum of the background counts was used as the background $b$ (see
Equation~\ref{e:poiprob} in \S\ref{s:countsul}) when finding the
combined counts upper limit, and the sum of the exposure times was
used as the combined exposure time when finding the combined flux
upper limit (see \S\ref{s:fluxul}).  Observed source and background
counts for all detectors are listed in Table \ref{t:xul}.  

\subsection{\rosat}
\label{s:data_rosat}

Bergh\"{o}fer et al.\ (1999) find an X-ray count rate upper limit of
$5\times10^{-4}$~ct~s$^{-1}$ for \bet\ based on \rosat/HRI observations,
and a limit of $10^{-2}$~ct~s$^{-1}$ based on the \rosat\ All-Sky Survey
data (RASS).
For completeness, we have included these estimates in
Figure~\ref{f:cecf_flx_bol_solar},
and have also recalculated the limit based on RASS data.
Because \bet\
is not detected, we estimate the upper limit (see \S\ref{s:ul})
based on the high-resolution X-ray background maps derived
from RASS (Snowden et al. 1997).
These maps are exposure corrected images with $12'\times12'$
pixels, in seven energy bands, each with a corresponding error map.
We considered bands 1 and 2, spanning $0.11-0.284$~keV, and bands 1 through
7, spanning $0.11-2.04$~keV.  By averaging the exposure times for eight
sources in the RASS Faint Sources Catalog within $1^{\circ}$ of \bet,
we estimated the exposure time for the local background to be $460\pm5$~s.
To estimate the counts upper limit, we collected background
counts within a circle with $r=35''$ centered on the position of
\bet; such a circle encloses 95\% of the PSPC PSF.  Observed source
and background counts are listed in Table \ref{t:xul}.  We obtain a limit
nearly identical to that derived by Bergh\"{o}fer et al.\ for the RASS.

\subsection{UV, Optical, and IR Leak}
\label{s:data_uvleak}

An important concern when optically bright sources are X-ray weak
is that counts that may be detected in the instrument could be
due to its so called out-of-band response to UV, optical, and IR
photons.  It is therefore critical to include this leak into
estimates of the background.  That is, we must compute the X-ray
detection limit that includes possible contributions to the observed
count rate from lower energy photons.

For the PSPC, we adopt a
conservative estimate of $\sim 2\times10^{-3}$ ct s$^{-1}$ for
the UV response as quoted in the \rosat\ User's
Handbook (expected for a high-temperature star such
as Vega).\footnote{http://wave.xray.mpe.mpg.de/rosat/doc/ruh/rosathandbook.html}
This implies that only 1 UV photon will be detected during a 460~s
exposure.  For \chandra, an analysis of the out-of-band response of the HRC
to \bet\ (Posson-Brown \& Kashyap 2005) shows that the UV leak
from \bet\ is vanishingly small (Table~\ref{t:uvtab}) and no UV
photons will be detected during the HRC observations.  A similar
estimate also applies to ACIS-I observations (Wolk 2002).

\begin{deluxetable}{ccccc}
\tabletypesize{\small}
\tablewidth{0pt}
\tablecaption{UV leak upper limits for HRC observations of \bet\ at the
  aimpoint.  The predicted count rates (and $1\sigma$ bounds) based on
  the UV/optical flux of \bet\ are compared with the upper limits
  derived from the accumulated data.  The time required for a
  detection at 99.7\% significance, derived from comparison of the
  predicted rate to the upper limit, is also noted. (Posson-Brown
  \& Kashyap 2005) \label{t:uvtab}}
\tablehead{
\colhead{Detector} & \colhead{Expected count rate} &
\colhead{Accumulated} & \colhead{Upper Limit} &
\colhead{Required}\\ 
\hfill & [\ctrt] & Exposure [s] & [\ctrt] & Exposure
}
\startdata
HRC-I & $1.29^{+0.11}_{-0.14} \times 10^{-6}$ & 7991 & $7.04 \times 10^{-4}$ & $\sim2\times10^{9}$ s \\
HRC-S & $1.64^{+0.86}_{-0.71} \times 10^{-5}$ & 7832 & $1.86 \times
10^{-3}$ & $\sim 9\times10^{7}$s \\
\enddata
\end{deluxetable}

\section{Analysis}
\label{s:ul}

\subsection{Counts Upper Limits}
\label{s:countsul}

To calculate the X-ray counts upper limits, we estimate the counts that
would need to be present in order to detect the source above a given
background, assuming a Poisson probability distribution for background
counts.  The probability that at most $D$ counts would be observed as a
statistical fluctuation, given a background $b$, is 
\begin{equation}
\label{e:poiprob}
p(\le D \mid b) = \sum\limits_{i=0}^D \frac{b^i e^{-b}}{i!} \,.
\end{equation}
The background $b$ is determined locally, from an annulus around
the nominal location of the source, and where applicable, an
estimate of the UV leak is added to it (see \S\ref{s:data}).
Note that this is a cumulative probability estimate, and uses
the full Poisson likelihood in the process. 
For a specific probability threshold $p$, the upper limit is 
$ul(p) = D(p) -b$. We allow for statistical variations in the
background counts by doing Monte Carlo simulations where the
background counts are sampled from a Poisson distribution.
We typically report a ``$3\sigma$'' limit, corresponding to a
probability of 0.997 that matches the integrated area under a normalized
Gaussian between $\pm3\sigma$.  Note that in our case, for
consistency with the Poisson distribution in the low-counts
regime, the integration always starts at a counts intensity of $0$,
and the counts value at which the probability threshold is met
does not coincide with the aforementioned $\pm3\sigma$ range of
a Gaussian.
For a more detailed description of this approach,
see Pease et al.\ (2006).
Our results are summarized in Table~\ref{t:xul}.

\begin{deluxetable}{lrrrrrc}
\tabletypesize{\small}
\tablecaption{$3\sigma$ Counts Upper Limits on \bet\ \label{t:xul}}
\tablewidth{0pt}
\tablehead{
\colhead{Detector} & \colhead{Exposure} & 
\multicolumn{2}{c}{Background Region} &
\multicolumn{2}{c}{Source Region} & \colhead{Upper Limit} \\
\hfill & $t_{\rm exp}$ & $a_{\rm bkg}$ & $N_{\rm bkg}$ 
& $a_{\rm src}$ & $N_{\rm src}$ & $D-b$ \\
\hfill & [s] & [arcsec$^{2}$] & [ct] & [arcsec$^{2}$] & [ct] & [ct]
}
\startdata
\rosat/PSPC bands 1-2 & 460   & 4147200 & 234   & 3848  & 38 & 6 \\
\rosat/PSPC bands 1-7 & 460   & 4147200 & 358   & 3848  & 48 & 6 \\
\\
{\cha}/HRC-I         & 7791  & 14126   & 1052  & 2.66  & 0  & 2 \\
{\cha}/HRC-S         & 7832  & 16245   & 9340  & 2.66  & 4  & 4 \\
{\cha}/ACIS-I        & 4897  & 31102   & 50    & 28.27 & 0  & 2 \\
{\cha} combined\tablenotemark{a}      & 20720 & 31102   & 20384 & 28.27 & 19 & 13 \\
\enddata
\tablenotetext{a}{Note that the size of the source region used when
  finding the combined limit is larger than the source sizes used when
  finding the HRC limits (see \S\ref{s:data_chandra})}
\end{deluxetable}

\subsection{Flux Upper Limits}
\label{s:fluxul}

The counts upper limit determined above in \S\ref{s:countsul}
represents the minimum counts that \bet\ must have in the source
aperture in order to be detected.  If it were observed with
the same instrumental sensitivity a large number of times,
and if it had a true flux intensity that produces, on average,
the same number of counts in the source aperture as the quoted
counts limit, then it would be detected in half the observations
and not detected in the other half.  A simple multiplicative
transformation of the counts limit to the flux limit can therefore
be performed under these conditions by computing a counts-to-energy
conversion factor for various putative plasma temperatures.

\begin{figure}[htb!]
\begin{center}
{\includegraphics{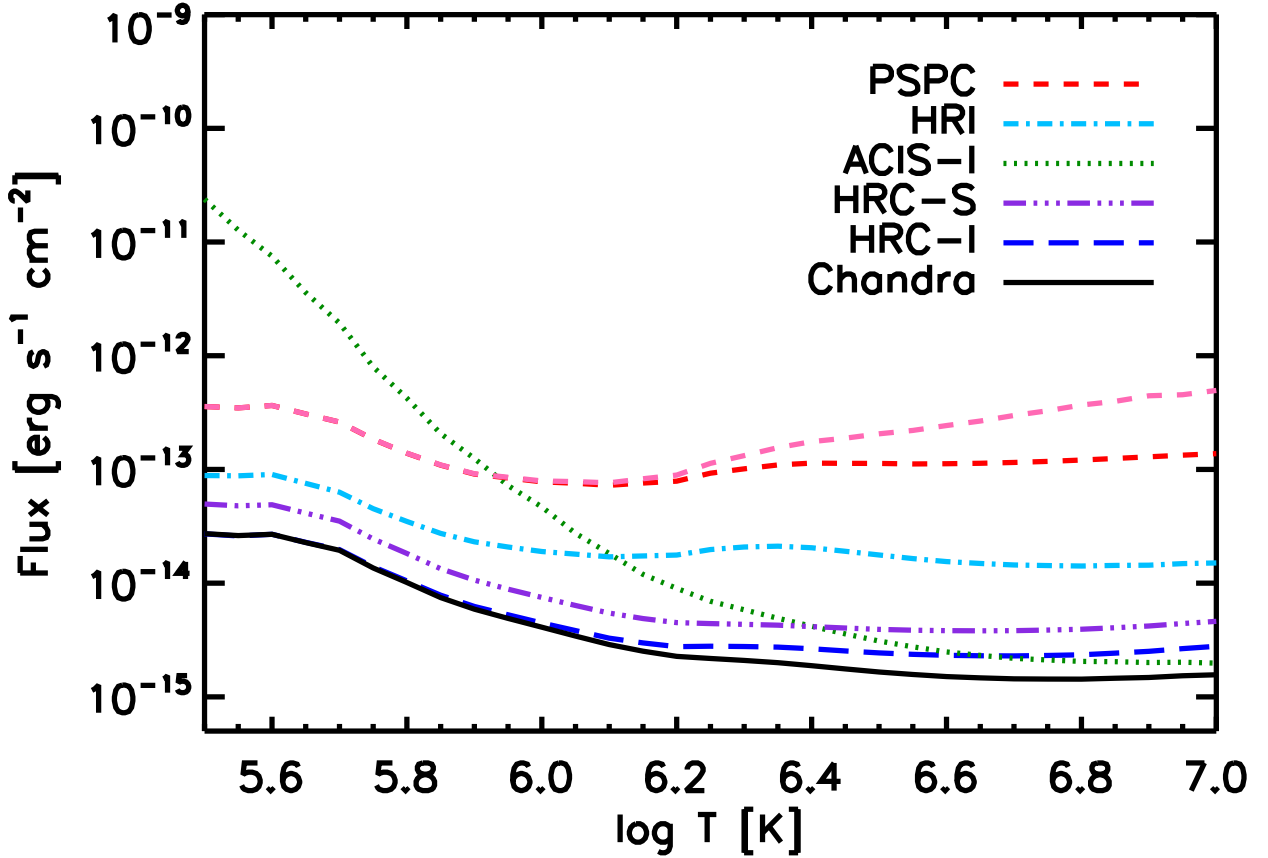}}
\caption{Upper limits on the observed X-ray flux from Betelgeuse
at Earth, as a function of coronal temperature, based on non-detections
from \chandra\ and \rosat.  The upper limit curves derived from each
of the instruments are shown (PSPC: dashed; we show the results for
both the full band [lower curve] and for just the soft band [channels
R1 and R2; upper curve];
HRI (from Bergh\"{o}fer et al.\ 1999): dot-dashed; ACIS-I: dotted; HRC-S: dash-triple-dotted; HRC-I: long-dashed)
as well as the limit from the combined \chandra\ observations (solid line).
Note that the \chandra\ limits are significantly below those from \rosat,
except for ACIS-I at low temperatures.  The combined limit from \chandra\
improves upon previous observations by roughly an order of magnitude.}
\label{f:cecf_flx_bol_solar}
\end{center}
\end{figure}

For each instrument, we generate spectra for isothermal
plasma at a given emission measure over a range of temperatures,
$T=0.3-10$~MK, using PINTofALE (Kashyap \& Drake 2000).  We
use the CHIANTI database of atomic emissivities (Dere et al.\ 1997, Young
et al.\ 2003) with solar abundances (Grevesse et al.\ 1992)
and ion balances from Mazzotta et al.\ (1998).
These spectra, which define the incident intensity at the
telescope, are multiplied by the appropriate effective
areas and convolved with the instrument response matrices
to obtain the counts predicted in a given passband.
The ratio of the incident flux to the derived counts gives a
counts-to-flux conversion factor for each temperature,
which then allows us to convert the counts upper
limits to flux upper limits.  These upper limits are shown
in Figure~\ref{f:cecf_flx_bol_solar}.  

In addition to considering the \chandra\ instruments separately, we also
combine them to achieve a more stringent upper limit on \bet's X-ray flux.
We do this by constructing a virtual observation that is essentially
the sum of the individual observations, weighted by the appropriate
effective areas and exposure times.  For a nominal incident flux
from a plasma at a given temperature,
we compute the counts predicted for the specific combinations of
\chandra\ instruments and exposures in Table~\ref{t:obsinfo}, and
compute the sum of these counts.  Because the relative fractions
of the counts remain unchanged as the nominal flux changes,
the ratio of the summed counts and the input flux gives the
counts-to-flux conversion factor for this virtual observation.
We thus obtain the most stringent
limits on the X-ray flux from \bet\ obtained thus far, which ranges
from $\approx2\times10^{-14}$~ergs~s$^{-1}$~cm$^{-2}$ at $T<0.5$~MK to as low
as $2.2\times10^{-15}$~ergs~s$^{-1}$~cm$^{-2}$ at $T=50$~MK.

\begin{figure}[htb!]
\begin{center}
{\includegraphics[width=6in]{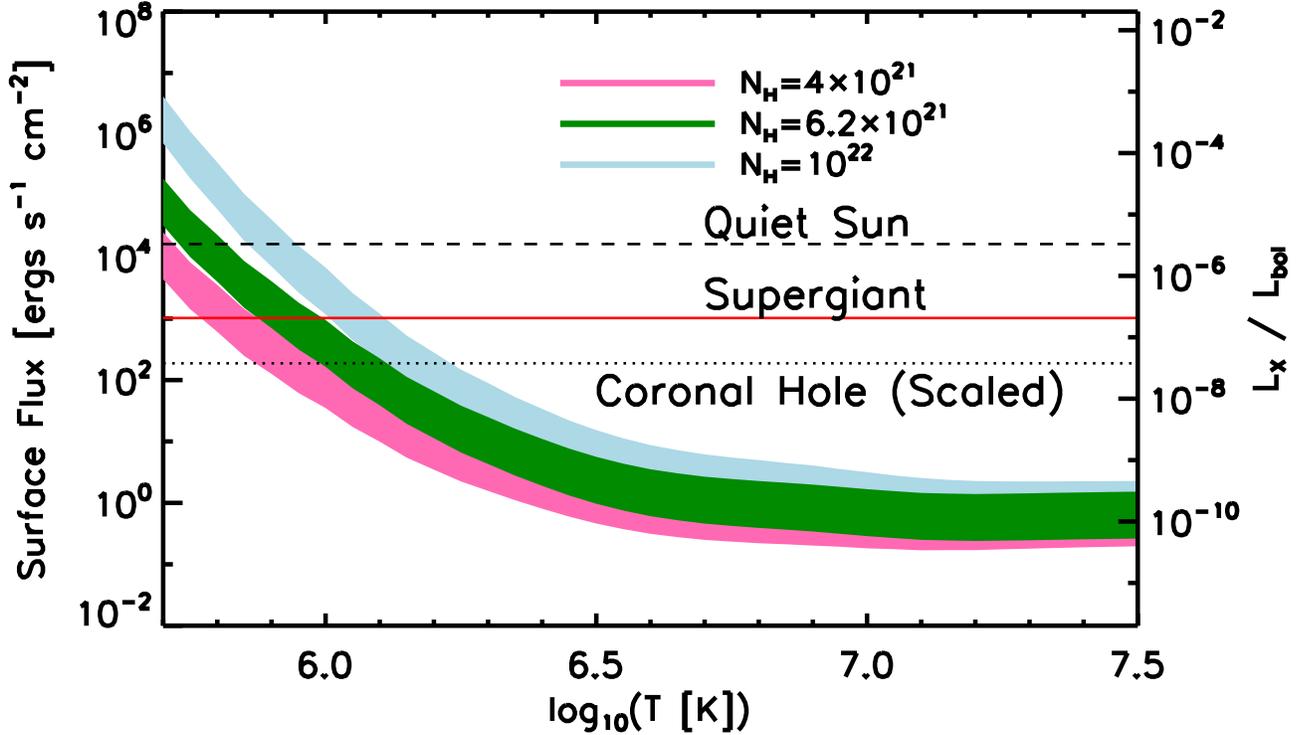}}
\caption{Upper limits on the surface X-ray flux, $\surffx$ of Betelgeuse as a
function of coronal temperature, based on non-detections from \chandra.
The $\surffx$ are calculated for three different values
of the column density, one for the nominal column,
$N_H=6.2\times10^{21}$~cm$^{-2}$ (dark curve in the middle)
and the other two representing the uncertainty in our
choice of $N_H$ (light curves), at $N_H=4\times10^{21}$
(lower curve) and $10^{22}$~cm$^{-2}$~cm$^{-2}$ (upper curve).
The width of the shaded regions represent the measurement errors
in the distance and diameter of \bet, as well as the statistical
uncertainty in the precision with which the counts upper limit is
determined.  The horizontal lines represent the surface flux of
the quiet Sun (dashed), the same flux scaled to match the magnetic
field strength on \bet\ (solid), and the flux from
solar coronal holes scaled to match \bet\ (dotted); this
is discussed in \S\ref{s:magnetic_carpet}.
The scale
along the right represents the $\lx/\lbol$ values corresponding
to the nominal distance and size of \bet.
\label{f:bet_ul_nH_err}}
\end{center}
\end{figure}

The flux limits at Earth can also be converted to intrinsic
X-ray luminosity limits at the star after a suitable absorption
column density $N_H$ is adopted.
The circumstellar shell surrounding \bet\ has a column density
estimated to be $N_H=6.2\times10^{21}$~cm$^{-2}$ (Hagen 1978),
and an additional column of approximately the same magnitude can be present due to chromospheric material that extends
above a conjectured ``buried corona'' (see Ayres et al.\ 2003).
We also consider a smaller value of $N_H$ since part of the
circumstellar column is likely to be in dust and grains, which
will effectively reduce the absorption column.
The ratio $\lx/\lbol$ and the corresponding average surface
X-ray flux $\surffx$ are shown in Figure~\ref{f:bet_ul_nH_err}.
The implications of the derived upper limits to the surface flux
are discussed in \S\ref{s:magnetic_carpet}.
Throughout this paper, we adopt
$\lbol=1.2\times10^{38}$~ergs~s$^{-1}$ for \bet, calculated from
the observed visual magnitude.

\section{Discussion}
\label{s:discuss}

\subsection{Magnetic Fields}
\label{s:magnetic_fields}

If any coronal plasma exists on the surface of \bet, it is
likely confined in place by magnetic fields.  From numerical
studies (see \S\ref{s:magnetic_carpet} below), we expect fields
of strength approaching 500~G.  How does it compare to the
observational limit?  Here we estimate the magnetic field
strength required to confine plasma, assuming that there
does exist coronal plasma emitting at a level just below the
derived flux upper limit (\S\ref{s:fluxul}), and further
assuming equipartition of
thermal and magnetic energy densities in the corona,
i.e.,
$\frac{B^2}{8\pi} = 2 n_e k_B T$, and hence
\begin{equation}
B = \left( \frac
{16 \pi \mu^2 k_B^2 T^2 \lx}
{h R_*^2 f \Lambda(T)}
\right)^{1/4}
\end{equation}
where $T$ is the assumed plasma temperature, $\Lambda(T)$ is
the power emitted from a unit volume of the plasma, $\mu$ is
the effective mass of an average particle ($\mu\approx\frac{1}{2}$
for fully ionized plasma), $R_*$ is the stellar radius, $k_B$
is Boltzmann's constant, $f$ is the filling fraction of
the X-ray active region on the surface of the star, and $h$ is
the height of the corona.  Note that $B~\propto~(\lx/f)^{1/4}$,
i.e., it is only weakly dependent on the X-ray luminosity and
the filling fraction.

\begin{figure}[htb!]
\begin{center}
{\includegraphics{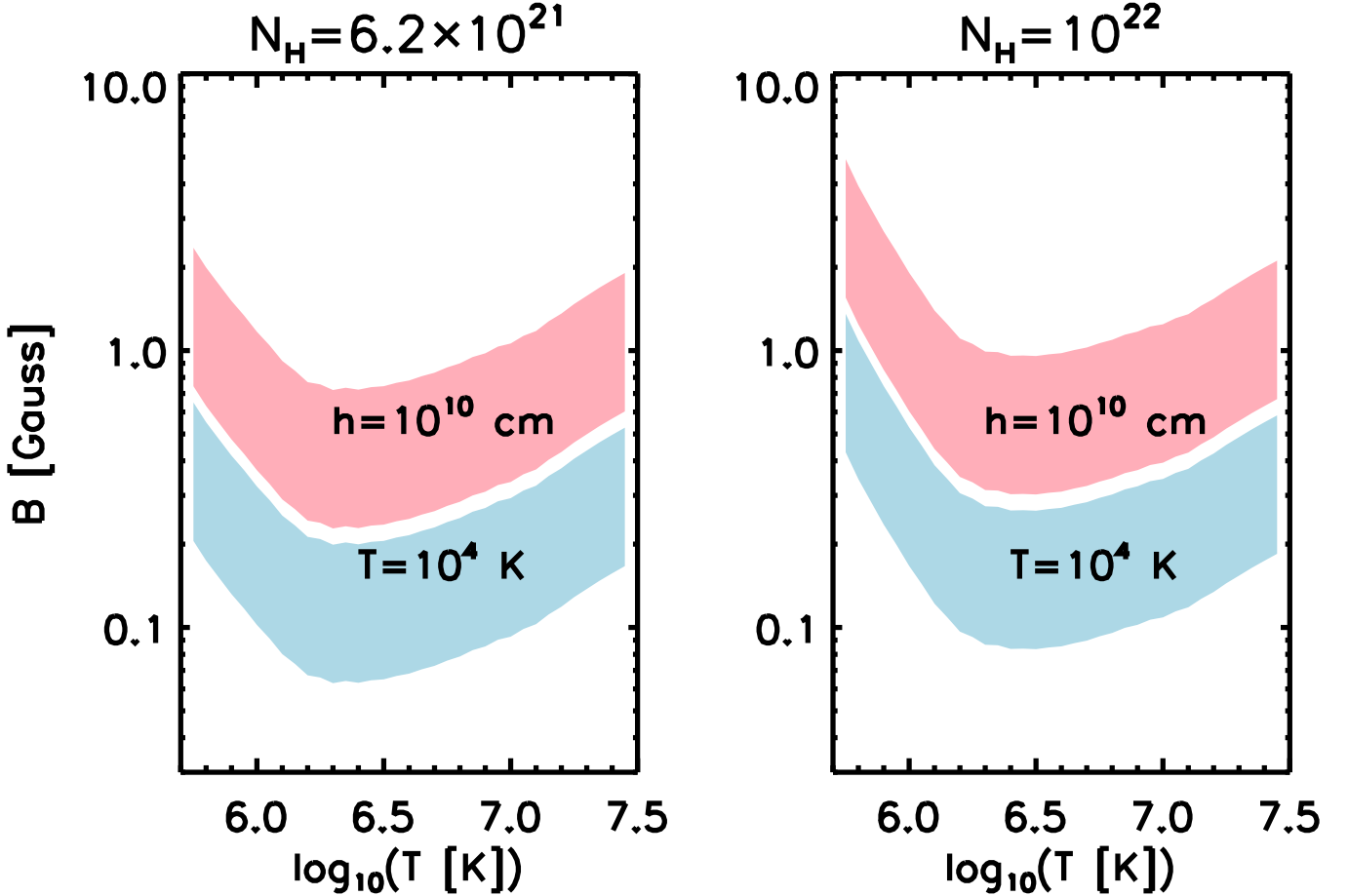}}
\caption{Magnetic field strength $B$ required to confine coronal plasma on \bet, as a function of the plasma
temperature.  The shaded regions represent the variation expected
due to a patchy coverage of the surface by the magnetic field,
with the lower edges corresponding to a filling fraction $f=1$
and the upper edges to $f=0.01$.  The lower shaded regions
represent a corona with the same height as the scale height of
the chromosphere, and for comparison, the upper shaded regions
represent a corona with a similar height as that of the solar
corona.  The two plots are calculated assuming different
column densities of absorption: $N_H=6.2\times10^{21}$ (left)
and $10^{22}$~cm$^{-2}$ (right).
\label{f:BTf} }
\end{center}
\end{figure}

In Figure~\ref{f:BTf}, we show the required field strength $B$,
for a range of filling fraction $f$, for different estimates
of the height of the corona.  We obtain estimates both for a
corona that is approximately the same height as the solar corona
($h=10^{10}$~cm) and for one that extends as high as the
chromospheric scale height
($h={kT_{\rm chrom}R_{*}^2}/{\mu~m_{\rm H}GM_{*}}\approx2\times10^{12}$~cm).
The latter is relevant in the context of a ``buried corona'' scenario
expounded by Ayres et al.\ (2003).
We find that the required $B$ ranges from $\lesssim0.1$~G up to $\approx5$~G,
considerably less than the field strength expected from modeling
(Dorch 2004).  Weaker fields are sufficient for lower levels of
X-ray emission.
(Note that our calculation of $B$ is not a limit,
and we cannot rule out the presence of a stronger magnetic field.
However, we can say that a stronger magnetic field is {\sl not required}.)
We thus conclude that sustaining a weak corona on \bet\ is
feasible, and such coronae cannot yet be ruled out based on
the X-ray flux limits.

\subsection{Magnetic Carpet}
\label{s:magnetic_carpet}

Numerical MHD simulations of stars such as \bet\ 
have shown that a highly structured magnetic dynamo may operate
on them (Dorch 2004, Freytag et al.\ 2002), with field strength
as high as $B_*\approx500$~G.
The energy spectra of the magnetic energy density peaks cascade
to small scales, with a preponderately high contribution at
scales $\sim \frac{1}{10}R_*$.  This is very similar to the
magnetic activity in the quiet Sun, where for comparison, field
strengths reach up to $\sim2000$~G (Cerde\~{n}a et al.\ 2006),
and the magnetic structures of mixed polarity emerge uniformly
over the surface on timescales approximately the same as the
convective cells.  This is the so-called `magnetic carpet'
(Schrijver et al.\ 1997, Title 2000) which pervades the surface
of the Sun away from active regions.

If red supergiants are magnetically active in a similar manner, 
the total energy dumped into the corona, and the resulting plasma
temperatures and X-ray luminosities, are similar in scope to that seen in the quiet Sun and in coronal holes,
scaled only by the energy densities and volumes involved.  
We assess the feasibility of this scenario by computing the
expected surface flux upper limit $\surffx$ for \bet.  Figure~\ref{f:bet_ul_nH_err} shows bands corresponding
to the surface flux limits, calculated assuming a variety of
plausible absorbing column densities,
$N_H=(4,6.2,10)\times10^{21}$~cm$^{-2}$ (see \S\ref{s:fluxul}).
Here we assume that the entire surface is covered by magnetic
structures that contribute to the X-ray emission, i.e., that
the filling fraction $f=1$.  However, supergiant stars like
\bet\ possess very large convection cells (Lim et al. 1998) which suggests
that $f << 1$.  Smaller values of $f$ imply that
the surface flux upper limit will be correspondingly higher.  We show
the effects of smaller filling fractions in Figure~\ref{f:BTf}.

The X-ray luminosity of the quiet Sun is $\approx10^{27}$~ergs~s$^{-1}$
(Golub \& Pasachoff 1997)
corresponding to a surface flux of
$1.64\times10^{4}$~ergs~s$^{-1}$~cm$^{-2}$.  If similar processes
to those on the Sun operate on \bet, it is reasonable to expect that
the energy flux deposited into the corona will be similar, scaled by the available magnetic energy density.  Thus, we expect
a surface X-ray flux of $\approx10^3$~ergs~s$^{-1}$~cm$^{-2}$ on
\bet.  Note however that the surface flux on \bet\ could be
lower if it were dominated by features similar to the solar
coronal holes.
The X-ray flux from
coronal holes on the Sun is much lower, $\approx3\times10^{3}$~ergs~s$^{-1}$~cm$^{-2}$
(Vernazza \& Smith 1977, Schrijver et al., 2004), and again scaling
it to the expected magnetic field energy density present on \bet, we obtain a
possible X-ray surface flux of $\approx200$~ergs~s$^{-1}$~cm$^{-2}$
(see Figure~\ref{f:bet_ul_nH_err}).

The upper limit on $\surffx$ is strongly dependent on the
temperature of the plasma because of the sensitive dependence
of the observable spectrum on the absorption column.  We
find that the limit is as low as $1$~ergs~s$^{-1}$~cm$^{-2}$
($\lx/\lbol \approx 10^{-10}$)
at high temperatures, but cannot be reduced below
$\approx10^2-10^3$~ergs~s$^{-1}$~cm$^{-2}$ for $T<1$~MK
($\lx/\lbol \approx 10^{-8}-10^{-7}$)
for the sensitivity achieved thus far with \chandra.  While the
upper limit at low temperatures still lies above the flux
expected from the solar analogy,
this calculation does rule out the existence of pervasive
quiet Sun type emission at high temperatures, since a surface flux
of $\approx10^2-10^3$~ergs~s$^{-1}$~cm$^{-2}$ from plasma at
$T>2$~MK would have been easily detected.  Note however that
this does not preclude patchy and highly localized regions of
magnetic activity that produces high temperature plasma, or
more pervasive low-temperature plasma emission arising from
this mechanism: plasma at $T\lesssim1$~MK will remain
undetected at the sensitivity limit of our observations.
Since the quiet Sun and coronal hole plasma is at $T\gtrsim1$~MK,
this further suggests that even if hot plasma at lower temperatures
is present on \bet, it will bear little resemblance to the solar case.

\subsection{Coronal Proxies}
\label{s:coronal_proxies}

Numerous UV and FUV chromospheric lines have been identified as
proxies for coronal activity in normal and giant stars.  Note
that the physical relationship between the mechanisms that generate
coronal and chromospheric line emission is poorly understood,
and the proxy lines are often formed at temperatures very
different from those that characterize coronae.  However,
assuming that similar processes occur on \bet\ as on coronally
active giants and main sequence stars, we investigate whether
our derived flux upper limits (see \S\ref{s:fluxul}) are consistent
with observations of coronal proxy lines on \bet.  Based on known
correlations between X-ray flux and proxy lines such as C\,IV,
Si\,IV, etc., we can estimate the X-ray luminosity that can be
expected from \bet.

For main sequence stars, Redfield et al.\ (2003) find a strong correlation
between the soft X-ray flux and the Fe\,XVIII\,$\lambda974$ flux
(see their Figure 7) from \fuse\
(Far Ultraviolet Spectroscopic Explorer) spectra of main sequence and late-type
stars.   Fe\,XVIII\, forms above $T=2$~MK, and has
a peak response at $T=6$~MK, and is thus sensitive to plasma temperatures similar to that on active binaries such as Capella.
Redfield et al.\ place an upper limit of
$8\times10^{-13}$~ergs~s$^{-1}$~cm$^{-2}$ on the Fe\,XVIII flux
from \bet\ ($L_{\rm FeXVIII}/\lbol=7.5\times10^{-11}$) which implies a value for the
unabsorbed X-ray luminosity $\lx/\lbol < 10^{-7}$ if supergiants such
as \bet\ follow the correlation seen in their  Figure 7.
The flux
upper limit we calculate (\S\ref{s:fluxul},
Figure~\ref{f:bet_ul_nH_err}) corresponds to
$\lx/\lbol<10^{-9}$
for $T>1$~MK, well below the value predicted by the main sequence correlation.
Conversely, assuming that the correlation is valid for supergiants,
our X-ray upper limit implies a stronger constraint on the Fe\,XVIII
emission, $L_{\rm FeXVIII}<7.5\times10^{-13}~\lbol$.

Similarly, there exists a clear correlation between
C\,IV and X-ray luminosities for giant stars (see, e.g., Ayres et al.\ 1997,
especially their Figure 2).  Even though there is evidence for
considerable scatter for different types of stars, one
can establish a general correspondence that is valid to
within an order of magnitude.  Based on \textit{IUE}
ultraviolet spectra, Basri et al.\ (1981) place a
$3\sigma$ upper limit of $1.5\times10^{-13}$~ergs~s$^{-1}$~cm$^{-2}$
on the C\,IV\,$\lambda1549$ flux from \bet, and this
translates to an unabsorbed X-ray luminosity limit
of $\lx/\lbol \lesssim 10^{-9}$.  In the temperature range
$T=1-10$~MK (which matches the sensitivity range of X-ray
measurements from {\sl Einstein} and \rosat, on which this
correlation is based), this limit is comparable
to our observational limit determined above.

Lastly, from \fuse\ spectra of \bet\ published by Dupree et
al.\ (2005; see their Figure 4), we estimate an upper limit
of $3\times10^{-14}$~ergs~s$^{-1}$~cm$^{-2}$ on its Si\,IV flux.
If the correlation found by Ayres et al. (2003, see their Figure 4)
for giant stars applies to supergiants like \bet, then the limit of
$f_{\rm Si\,IV} / f_{\rm bol} < 5\times10^{-10}$ corresponds to an
expected X-ray unabsorbed flux of $\fx / f_{\rm bol} \lesssim 10^{-9}$
which is again comparable to the observed limit we obtain.

Note that previous studies of \bet\ in the optical and UV have
seen no evidence for chromospheric temperatures above 6000 K
(e.g. Lobel et al. 2000 \& 2001 and Carpenter et al. 1994) and our
observations are consistent with those findings.

From the comparisons above, we conclude that we have now achieved
a sensitivity in the X-ray regime that is comparable to the
sensitivity achieved in the far UV with coronal proxy lines,
especially at temperatures $T>3$~MK. In combination with the
surface flux limit arguments
made above (\S\ref{s:magnetic_carpet}) and the stringent upper
limit on $\lx/\lbol$ obtained here, we conclude that any
high-temperature plasma will have to arise from a mechanism
other than that which normally operates on the Sun.

Finally, note also that
if the hot plasma is ``buried'' in the highly extended
chromospheric material (as suggested by Ayres et al.\ 2003),
then both the chromospheric proxies and the coronal X-ray
flux will be subject to significant absorption, and our
placement of \bet\ on these flux-flux correlation diagrams
will be systematically low.  In such a case, the limit
on $\lx/\lbol$ will be less restrictive, but will not affect
our conclusions.

\section{Summary}
\label{s:summary}

We have carefully analyzed over 20~ks of \chandra\ observations
of \bet\ in an effort to detect X-ray emission from the massive
red supergiant.  However, \bet\ remains undetected, and we derive an upper limit to the X-ray count rate by calculating
the rate that would have resulted in a detection given the
extant background.  We have converted this count rate limit to
a flux limit at the telescope by computing the response of
\rosat\ and \chandra\ instruments to isothermal plasma producing
optically-thin thermal emission, and thereby derive the most
stringent upper limits to the X-ray flux from \bet\ obtained
thus far.  We find a limit for the flux from \bet\ at the telescope of
$\fx < 4\times10^{-15}$~ergs~s$^{-1}$~cm$^{-2}$ for temperatures
$T>1$~MK.  At lower temperatures, we place a limit of
$\fx < 3\times10^{-14}$~ergs~s$^{-1}$~cm$^{-2}$ on the flux.

The flux limit at Earth can be converted to a stellar surface flux upper
limit and to an $\lx/\lbol$ limit using the known distance and
size of \bet.  We compare the surface flux limits with the flux
expected from a solar like emission mechanism, where a pervasive
magnetic field maintains a low-level corona, as in the quiet Sun
or solar coronal holes.  We rule out such emission at
temperatures $>1$~MK, but such emission is still feasible
at lower temperatures.  The minimum magnetic field necessary to
maintain such a corona is $<10$~G, well within theoretical
expectations.

We compare the $\lx/\lbol$ upper limit we derive with the limits
obtained from non-detections of coronal tracer lines such as
C\,IV, Si\,IV, and Fe\,XVIII\,$\lambda974$ and find that we
achieve sensitivities in the X-ray comparable to that in the
coronal proxies.  These limits reinforce the conclusions arrived
at above, that high-temperature plasma, even at levels expected in
the presence of stellar coronal holes, is absent
on \bet, but the existence of low-temperature plasma cannot
be ruled out.

\acknowledgments
JPB, VLK, DOP, and JJD were supported by NASA contract NAS8-39073 
to the \cxc\ during the course of this research.
We thank Tom Ayres for useful feedback on determining upper limits.
VLK thanks Andreas Zezas, Steve Saar, and Mark Weber
for useful discussions.
This research has made use of the SIMBAD database,
operated at CDS, Strasbourg, France.

\clearpage

\end{document}